\input{aipcheck}

\documentclass[
    ,final            
  ]
  {aipproc}

\layoutstyle{6x9}

\usepackage[utf8]{inputenc}
\usepackage{amssymb}

\begin{document}

\title{IceCube: Neutrino Messages from GRBs}


\author{A. Kappes\footnote{associated with the University of
    Wisconsin-Madison} ~for the IceCube Collaboration}{
  address={Friedrich-Alexander-Universität Erlangen-Nürnberg, Erlangen
    Centre for Astroparticle Physics, Erwin-Rommel-Str. 1, 91058
    Erlangen}
}

\begin{abstract}
  The mystery of where and how Nature accelerates the cosmic rays is
  still unresolved a century after their discovery. Gamma ray bursts
  (GRBs) have been proposed as one of the more plausible sources of
  extragalactic cosmic rays. A positive observation of neutrinos in
  coincidence with a GRB would identify these objects as sources of
  the highest-energy cosmic rays and provide invaluable information
  about the processes occurring inside these phenomena. Calculations
  show that a kilometer-scale neutrino telescope is necessary for this
  task. The idea of such a detector is now becoming reality as IceCube
  at the South Pole nears completion. The contribution reviews the
  status of the construction and operation of IceCube and summarize
  the results from searches for neutrinos from GRBs and similar
  phenomena with IceCube and its predecessor, AMANDA. At the end, an
  outline of future plans and perspectives for IceCube is given.
\end{abstract}

\maketitle


\section{Introduction}
Gamma-ray bursts (GRBs) are among the most violent events in the
universe and among the few plausible candidates for sources of the
ultra-high energy cosmic rays. So-called long-duration GRBs ($\gtrsim
2$\,s) are thought to originate from the collapse of a massive star
into a black hole \citep{apj:405:273}, whereas short-duration GRBs
($\lesssim 2$\,s) are believed to be the result of the merger of two
compact objects (e.g., neutron stars) into a black hole
\citep{nat:340:126}. Though quite different in nature both scenarios
are consistent with the currently leading model for GRBs, the fireball
model \citep{apj:405:278}, with the energy source being the rapid
accretion of a large mass onto the newly formed black hole. In this
model, a highly relativistic outflow (fireball) dissipates its energy
via synchrotron or inverse Compton radiation of electrons accelerated
in internal shock fronts. This radiation in the keV--MeV range is
observed as the gamma-ray signal. In case of long GRBs the energy
in gamma rays is typically of ${\cal
  O}(10^{51}$--$10^{54}$\,erg$\,\times\,\Omega / 4\pi)$ where $\Omega$
is the opening angle for the gamma-ray emission. Short GRBs are
observed to release about a factor 100 less energy.

In addition to electrons, protons are thought to be accelerated via
the Fermi mechanism, resulting in an $E^{-2}$ power law spectrum with
energies up to $10^{20}$\,eV \citep{prl:75:386,apj:453:883}. Protons
of ${\cal O}(10^{15}\,\mathrm{eV})$ interact with the keV--MeV photons
forming a $\Delta^+$ resonance which decays into pions
\citep{prl:78:2292}. In the decay of the charged pions, neutrinos of
energy ${\cal O}(10^{14}\,\mathrm{eV})$ are produced with the
approximate ratios ($\nu_e$:$\nu_\mu$:$\nu_\tau$) = (1:2:0)
\footnote{Here and throughout the rest of the paper $\nu$ denotes both
  neutrinos and antineutrinos.}, changing to about (1:1:1) at Earth
due to oscillations \citep{app:3:267}.  First calculations of this
prompt neutrino flux \citep{prl:78:2292,apj:521:928} used average GRB
parameters and the GRB rate measured by BATSE to determine an all-sky
neutrino flux from the GRB population.

In a similar way, so-called precursor neutrinos can be generated when
the expanding fireball is still inside the progenitor star
\citep{pr:d68:083001}. In this case, the accelerated protons interact
with matter of the progenitor star or synchrotron photons. However,
due to the large optical depth the synchrotron photons cannot escape
the fireball and, hence, no gamma-ray signal is observed. 

\section{The IceCube Neutrino Observatory}
IceCube \citep{app:26:155}, the successor of the AMANDA-II experiment
and the first next-generation neutrino telescope, is currently being
installed in the deep ice at the geographic South Pole. Its final
configuration will instrument a volume of about $1\,\mathrm{km}^3$ of
clear ice in depths between 1450\,m and 2450\,m. Neutrinos are
reconstructed by detecting the Cherenkov light from charged secondary
particles, which are produced in interactions of the neutrinos with
the nuclei in the ice or the bedrock below. The optical sensors, known
as Digital Optical Modules (DOMs), consist of a 25\,cm Hamamatsu
photomultiplier tube (PMT) housed in a pressure-resistant glass sphere
and associated electronics \citep{nim:a601:294}. They are mounted on
vertical strings where each string carries 60 DOMs. The final detector
will contain 86 such strings spaced horizontally at approximately
125\,m intervals\footnote{Six of these strings will make up a dense
  subarray in the clearest ice known as Deep Core, extending the
  sensitivity of IceCube at lower energies.}. Physics data taking with
IceCube started in 2006 with 9 strings installed. Currently, 79
strings are deployed. The completion of the detector construction is
planned for the year 2011.

\section{Searches for Neutrino Emission from GRBs}
\subsection{Detection channels}
Charged secondary particles generated in neutrino interactions produce
Cherenkov light patterns in the detector that can be separated into
two classes (\emph{channels}).

In case of an incoming muon neutrino which interacts via a charged
current reaction, the so-called \emph{muon channel}, the resulting
outgoing muon leaves a track-like pattern of PMT signals in the
detector. As the muon can travel up to several kilometer in water or
rock the interaction can happen far away from the detector. Due to the
long lever arm that is associated with the track-like pattern of PMT
signals, this channel provides the best angular resolution for the
direction reconstruction of the neutrino. For energies above 1\,TeV it
is better than $1^\circ$. At energies below $\sim 10$\,TeV it is
dominated by the unknown angle between the reconstructed muon and the
original neutrino. The angular resolution improves with energy and at
energies above $\sim 10$\,TeV it is dominated by the precision of the
direction reconstruction of the muon. The energy resolution in this
channel is rather poor ($\Delta \log E_\mu \approx 0.3$
\cite{proc:icrc07:zornoza:1}) as only part of the muon's energy is
deposited inside the sensitive detector volume. In addition, an
unknown amount of energy is transferred into the shower at the
neutrino interaction vertex.

The so-called \emph{cascade channel} consists of neutrino interactions
which happen inside or near the detector and do not produce a
high-energy muon. It is characterized by the Cherenkov light emitted
from charged particles in the generated shower, which, within the
position resolution of the detector, is point-like. Due to the short
scattering length in ice, the angular profile of the light is
isotropic after a few ten meters. As a consequence, the directional
information is almost completely lost and the angular resolution for
the neutrino direction is of the order of $30^\circ$.  On the other
hand, showers deposit all of their energy near the interaction vertex
allowing for a better energy reconstruction than in the case of muons.
In particular, in charged current electron neutrino interactions, all
of the neutrino's energy is transferred into showers, allowing for an
improved neutrino energy reconstruction.  In contrast to the muon
channel to which only muon neutrinos in charged current reactions
contribute, the cascade channel is sensitive to all neutrino flavors.

Searches for neutrinos from GRBs are performed in both channels, where
the main channel is the muon channel due to its superior angular
resolution.

\subsection{Analysis methods}
Searches for neutrinos from GRBs are performed with and without the
usage of information from other detectors, mostly satellites (e.g.,
Swift, Fermi), which trigger on the prompt gamma-ray emission of GRBs.
In this so-called \emph{triggered searches}, the knowledge of the
direction and time of the emission allows for a significant reduction
of background without loss of signal. This leads to an increase in
sensitivity to a level where a single detected neutrino from a GRB can
be significant at the $5\sigma$ level. Analyses are performed by
estimating the expected background during the GRB emission from
off-time data thereby avoiding uncertainties in the detector
simulation for the calculation of the significance of a potential
signal.

On the other hand, triggered searches rely on the detection of GRBs by
satellites and hence miss all GRBs that are not seen because either
they are not in the field of view of satellites or because the GRBs do
not emit gamma rays. An example for the latter is the potentially
large class of choked GRBs where the jet is not powerful enough to
reach the surface of the progenitor star. In these cases, like in the
precursor phase, neutrinos and gamma rays are emitted but only
neutrinos can escape the dense source. Hence, the only detectors that
can observe these GRBs are neutrino telescopes. In order to find a
signal from these sources in the IceCube data, so-called \emph{rolling
  searches} are performed. In these analyses, a time window with a
width fitting the expected neutrino emission duration (typically
between 1\,s and 100\,s) is moved over the detected events comparing
the observed number of events with those expected for pure background.
Though these searches are sensitive to all GRBs with neutrino
emission, they are less sensitive for known GRBs due to the large
trial factor resulting from the sliding window.

\begin{figure}[t]
  \includegraphics[width=.55\textwidth]{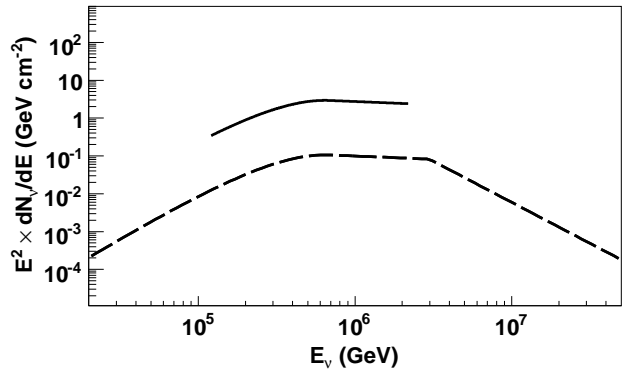}
  \caption{Calculated neutrino spectrum for GRB\,080319B ($\Gamma =
    300$; dashed line) and 90\% CL upper limit from the analysis of
    IceCube data taken with 9 strings (solid line). Taken from
    \cite{apj:701:1721}.}
  \label{fig:grb080319b}
\end{figure}

\subsection{Detection of Individual GRBs}
According to current calculations, the expected number of detected
events from an average GRB is rather low even in a km$^3$-size
neutrino telescope like IceCube (for an average Waxman-Bahcall burst
it is of the order of 0.1 events). However, as discussed in
\cite{app:20:429} the burst parameters vary significantly from burst
to burst leading to a large variation in the expected number of
detected events. Thus, the individual analysis of data from
exceptionally bright GRBs is highly interesting.

Such an analysis was previously performed with data from the AMANDA-II
detector \cite{proc:icrc05:stamatikos:1} with negative results. During
IceCube operations, the so-called ``naked-eye'' GRB\,080319B occurred
which was the optically brightest GRB ever recorded. At that time,
IceCube was running in a 9-string configuration. Calculations yielded
that the expected number of detected events for a jet bulk Lorentz
factor of 300 would be about 0.1 \cite{apj:701:1721}. The analysis of
the data \cite{apj:701:1721} showed no significant excess above the
background leading to the 90\% CL upper limit on the neutrino flux
shown in Fig.~\ref{fig:grb080319b}.

For the 10 times larger completed IceCube neutrino telescope, similar
bright GRBs are expected to yield of the order of 1 event in the
detector. Hence, the analysis of individual bright GRBs remains
interesting.

\begin{figure}
  \includegraphics[width=.55\textwidth]{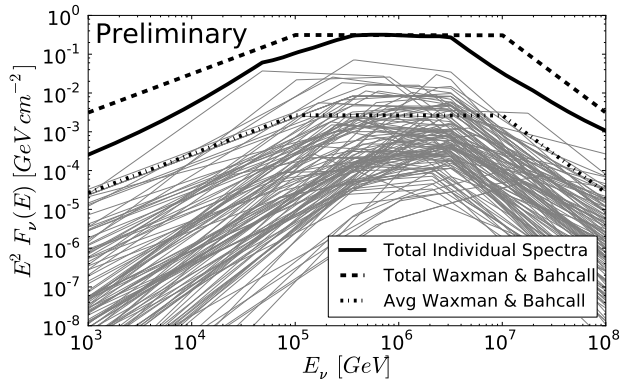}
  \caption{Calculated neutrino spectra for 117 bursts for which data
    has been recorded with the IceCube 40-string configuration (thin
    lines). Also shown is the sum of the 117 individual spectra (thick
    solid line), the flux from an average Waxman-Bahcall burst
    (dash-dotted line) and the sum of 117 average Waxman-Bahcall
    bursts (dashed line).}
  \label{fig:grbSpectra}
\end{figure}
\begin{figure}
  \includegraphics[width=.4\textwidth]{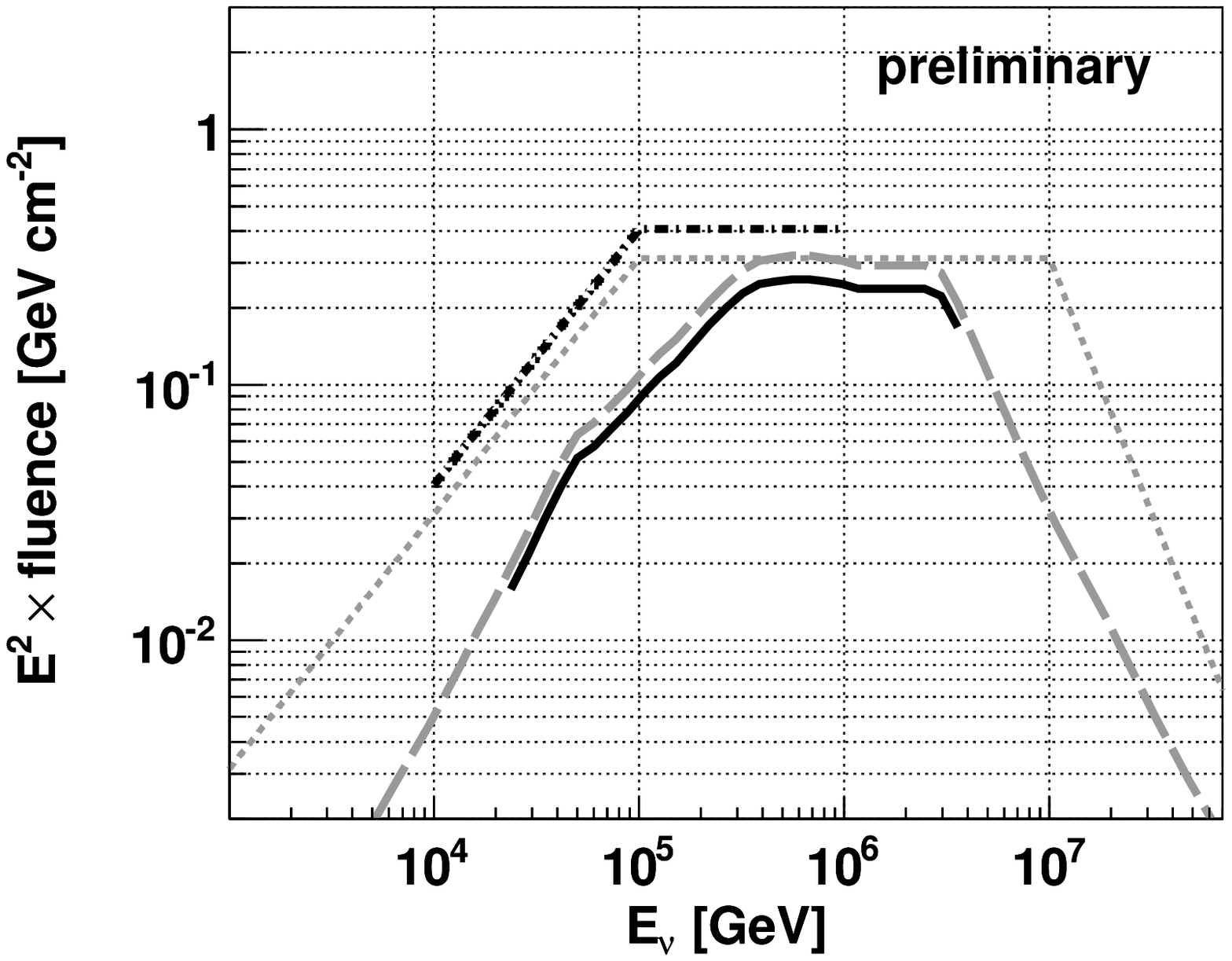}\hspace{0.1\textwidth}
  \includegraphics[width=.4\textwidth]{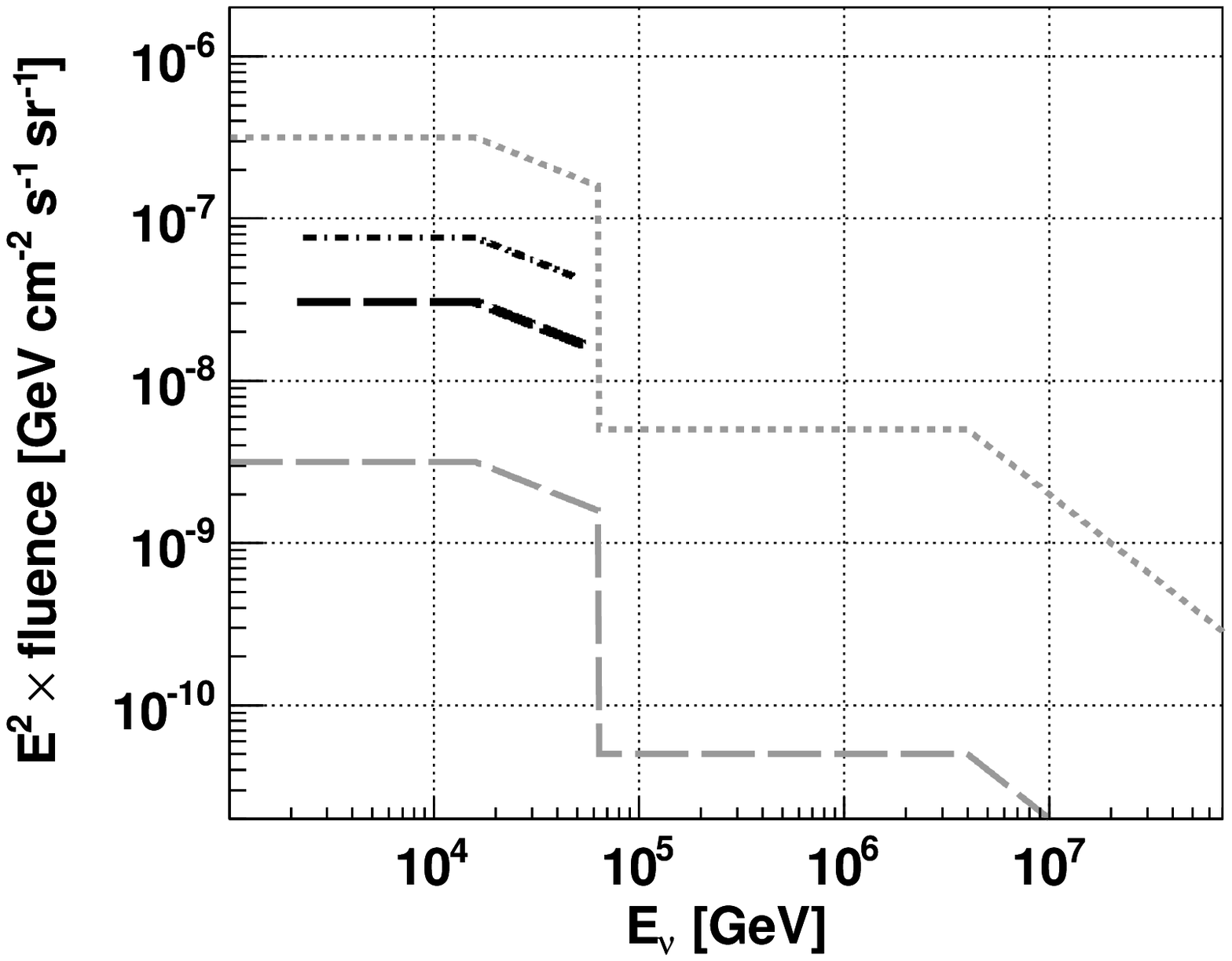}
  \caption{Left: Calculated total prompt neutrino spectrum from 117
    bursts during the operation of IceCube with 40 strings (light
    solid line) and 90\% CL upper limit (dark dashed line). In
    addition, the flux and 90\% CL upper limit from the analysis of
    416 bursts with the AMANDA-II detector (dash-dotted dark and light
    dotted line; scaled to 117 bursts) \cite{apj:674:357} is
    displayed.  Right: Flux of precursor neutrinos according to
    \cite{pr:d68:083001} (light dashed line) and the flux scaled by a
    factor 100 (light dotted line) which assumes that all type II SNe
    exhibit choked jets. The dark lines represent the 90\% upper
    limits from an analysis of IceCube 22-string data
    \cite{apj:710:346} (dark dashed line) and a rolling search in the
    cascade channel with AMANDA-II data \cite{apj:664:397} (dark
    dash-dotted line).}
  \label{fig:limits}
\end{figure}

\subsection{Stacked GRB Searches}
As the mean number of expected neutrinos from individual GRBs is
usually small, stacking the events from the directions of several GRBs
enhances the chances for a discovery. The disadvantage is that a
potential signal cannot by implication be identified with a neutrino
flux from a specific GRB.

So far, the most sensitive analysis with such an approach has been
performed with data taken with IceCube in its 40-string configuration
in 2008/09. Figure~\ref{fig:grbSpectra} shows the calculated neutrino
spectra for 117 bursts (mainly detected by Swift and Fermi) that
occurred during data taking. Differences to the flux from an average
Waxman-Bahcall burst, which is also shown, are clearly visible. They
are due to the fact that the average parameters of the GRBs in the
BATSE sample used by Waxman and Bahcall for their calculations
\cite{prl:78:2292} differ from those in the Swift/Fermi sample (see
\cite{apj:710:346} on how individual spectra are calculated). The
analysis of the data yielded no excess above the background. The left
plot in Figure~\ref{fig:limits} shows the resulting upper limits
together with the calculated total neutrino flux. The properly scaled
limit from the analysis of 416 bursts with the AMANDA-II detector
\cite{apj:674:357} is also shown. Though only half its final size and
using about a factor 4 fewer bursts, IceCube is already more sensitive
than AMANDA-II and starts to set limits below the model predictions
for prompt neutrino emission.

The right plot in Fig.~\ref{fig:limits} shows the current best
limits on the neutrino flux from the precursor phase. The dashed thick
line results from an analysis of data from 41 bursts taken with
IceCube in a 22-string configuration \cite{apj:710:346}. The thin
dash-dotted line is the limit from a rolling search in the cascade
channel with the AMANDA-II detector \cite{apj:664:397}. Though less
sensitive than the IceCube analysis the latter is currently the best
limit on neutrino fluxes from GRBs not detected in gamma rays. The
limit excludes the case where all type II supernovae (SNe) exhibit
choked jets (light dotted line).

An alternative approach to the detection of neutrinos from
core-collapse SNe will be discussed in the next section.

\begin{figure}
  \includegraphics[width=.55\textwidth]{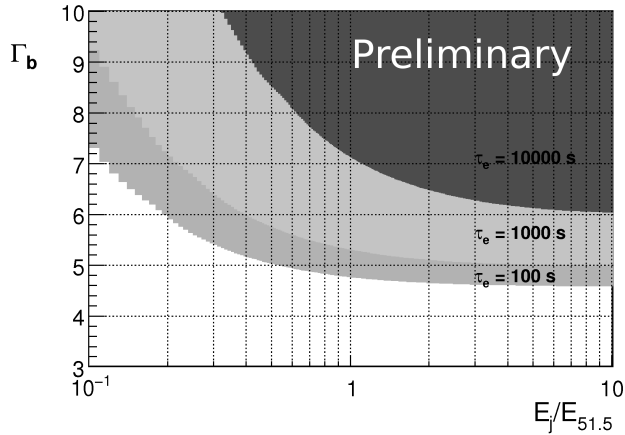}
  \caption{90\% CL upper limits in the $\Gamma_b$ (bulk Lorentz factor
    of the jet) and $E_j/E_{51.5}$ (jet energy divided by
    $10^{51.5}$\,erg) plane for three different search windows with
    width from 100\,s to 10\,000\,s. Derived from the non-observation
    of neutrinos from SN\,2008D under the assumption that its
    hypothesized jet was pointing towards Earth.}
  \label{fig:ccSN}
\end{figure}
\subsection{Neutrinos from Core-Collapse Supernovae}
Optical telescopes detect several hundred core-collapse SNe per year.
The SN light curves rise during the first few days after which a slow
decay over typically several tenths of days starts. The usual
precision of the measurement allows the determination of the actual
time of explosion to about one day \cite{app:33:19}, much larger than
the expected neutrino emission time. This results in a large
background when searching for associated neutrino emission and hence a
decreased sensitivity.

In January 2008, the X-ray telescope aboard the Swift satellite was
conducting a routine observation of NGC\,2770 when it recorded a
bright X-ray flash \cite{na:453:469} that was later associated with
SN\,2008D, a core-collapse SNe of type Ib. The X-ray flash was
interpreted as the signature of the shock break out. It provides the
most precise information on the time of the SN explosion to date and
hence allows for small neutrino search windows. Based on the model in
\cite{prl:95:061103} which predicts neutrino emission aligned with
hypothesized jets, the data was analyzed using three search windows
with durations between 100\,s and 10\,000\,s. No event was observed in
any of the three search windows and upper limits in the
$\Gamma_b$-$E_j$ parameter plane were determined (see
Fig.~\ref{fig:ccSN}).

\subsection{The Optical Follow-Up Program}
Often data from other detectors are used to guide analyses of IceCube
data. On the other hand, a coincidence of several neutrinos from a
given direction within a short time window is itself an interesting
target for observations with optical telescopes. Possible scenarios
are a rising SN light curve over several days or the decreasing
intensity of a GRB afterglow.

\begin{figure}
  \includegraphics[width=.5\textwidth]{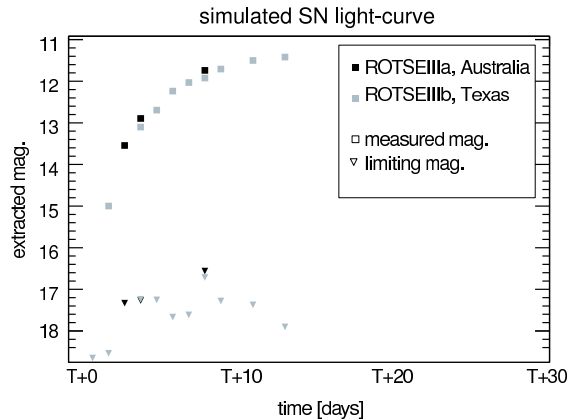}
  \caption{Results of the extraction of a simulated SN light
    curve which was injected into real background images taken with
    two of the ROTSE telescopes. The squares represent the measured
    magnitude. The limiting magnitudes are displayed as triangles.}
  \label{fig:snLightcurve}
\end{figure}

This so-called \emph{optical follow-up} program of IceCube
\cite{app:27:533} searches for two or more neutrinos within a angular
window of $4^\circ$ which occur within 100\,s. In case of a signal,
a trigger is sent with less than 5 minutes delay (in 2009, before an
upgrade to the processing system, the latency was a few hours) to
robotic telescopes which perform a follow-up program during the
following 14 nights.  Currently, the program includes the four
telescopes of the ROTSE network which have a large field of view of
$1.85^\circ \times 1.85^\circ$ comparable to the angular resolution of
IceCube and an almost 24 hour night-sky coverage. The images taken are
automatically processed and variable sources are extracted. The
results of a simulation using real background data and injected SN
light curves is shown in Fig.~\ref{fig:snLightcurve}.  The rising part
of the SN light curve is clearly visible.  The system is successfully
running since end of 2008 and data analysis is currently underway.

\section{Future Perspectives with IceCube}
Future IceCube analyses will continue to span a wide range of
scenarios from the individual analysis of exceptionally bright bursts
over the stacked analysis of a large number of bursts detected by
satellites to rolling window searches. For triggered searches, a good
and high sensitivity coverage of the sky with gamma-ray satellites is
essential. The Swift and Fermi satellites will operate at least until
2014 and 2013, respectively, where a 5 year extension for Fermi is
likely. Future satellites like SVOM (start planned in 2012), UFFO
(start planned in 2015) and EXIST (possible start in 2017) will
complement the existing ones and will provide a good coverage at least
until 2020.

The fact that GRBs are still not very well understood requires a
double-track approach. Model-specific analyses yield a high
sensitivity as they are optimized for the corresponding neutrino
spectra. On the other hand, in order not to miss any unknown mechanism
of neutrino production more generalized but less sensitive searches
that cover a large time window around GRBs are being performed and
remain mandatory.

The sensitivity of the IceCube detector is still increasing
significantly during the next years. Its operation is foreseen for at
least 10 years. Given the fact that already with only half of its
strings installed and one year of observation time it starts to set
limits below flux predictions, the chances for neutrino detection from
GRBs in the upcoming years are good. With a sensitivity about 10 times
larger than needed to detect the Waxman-Bahcall GRB flux, generic
models that assume GRBs as the major sources of ultra-high energy
cosmic rays will be tested.


\begin{theacknowledgments}
  The author thanks the organizers for the invitation to this
  interesting workshop. He gratefully acknowledges the support by the
  EU Marie Curie OIF Program for this work.

\end{theacknowledgments}

\end{document}